\documentclass[sigplan, 10pt]{acmart}

\usepackage[english]{babel}
\usepackage{blindtext}
\usepackage{graphicx}
\usepackage[font=small,labelfont=bf]{caption}
\usepackage{subcaption}
\usepackage{xr}
\usepackage{easy-todo}
\usepackage{array}
\usepackage{enumitem}
\newcolumntype{?}{!{\vrule width 1pt}}

\setlist{nosep, before=\vspace{0.3\baselineskip}, after=\vspace{0.3\baselineskip}}

\renewcommand\footnotetextcopyrightpermission[1]{} 
\setcopyright{none}

\acmDOI{}

\acmISBN{}

\acmConference[Submitted for review]{}
\acmYear{2021}

\acmPrice{}

\begin{document}
\title{Spindle: Techniques for Optimizing Atomic Multicast on RDMA}


\author{
{\rm Sagar Jha$^1$, Lorenzo Rosa$^{1,2}$ and Ken Birman$^1$}\\ {\it $^1$Cornell University, $^2$University of Bologna}
}

\renewcommand{\shortauthors}{Jha, Rosa and Birman}

\begin{abstract}
Leveraging one-sided RDMA for applications that replicate small data objects can be surprisingly difficult: such uses amplify any protocol overheads. Spindle is a set of optimization techniques for systematically tackling this class of challenges for atomic multicast over RDMA. These include memory polling optimizations using novel sender and receiver batching techniques, null-message send logic, and improved multi-thread synchronization. We applied Spindle to Derecho, an open-source C++ library for atomic multicast, and obtained significant performance improvements both for the library itself and for an OMG-compliant avionics DDS built over Derecho.  Derecho's multicast bandwidth utilization for 10KB messages rose from 1GB/s to 9.7GB/s on a 12.5GB/s network, and it became more robust to delays.  Interestingly, although some of our techniques employ batching, latency dropped by nearly two orders of magnitude.  Spindle optimizations should also be of value in other RDMA applications limited by the speed of coordination.
\end{abstract}

\maketitle

\section{Introduction}
\label{sec:intro}
\footnote{This work has been submitted to the IEEE for possible publication. Copyright may be transferred without notice, after which this version may no longer be accessible.}
In settings such as avionics,  developers are required to adhere to standards like the OMG DDS ~\cite{OMG-DDS},  a form of publish-subscribe using 8-bit topic numbers and messages represented as byte vectors.  Avionics DDS applications include both safety-critical flight-management systems for aircraft control and less critical onboard  infrastructures that support higher-level applications, and hence require various levels of DDS quality-of-service.

Because the DDS API  is platform-independent and easy to implement, a natural way to leverage modern hardware such as RDMA is to layer the DDS API over an RDMA-capable communications library that offers suitable functionality.  We selected a mature open-source library, Derecho~\cite{derecho-tocs}, which
offers point-to-point and multicast communication options with a range of QoS guarantees (failure atomicity, total ordering, message logging with durability) \cite{derecho-tocs}.  
Derecho supports both fast datacenter TCP and RDMA; in the latter mode, it sends small messages using one-sided RDMA writes ~\cite{rdma-manual}.  In this model, a receiver designates a region of memory into which the sender is permitted to write at will, then polls for updates. RDMA writes bypass the kernel, avoid excessive coordination, and achieve a zero-copy critical path with high throughput (> 200 Gbps) and low latency (\textasciitilde 1-2$\mu s$).   

Our work revealed that although Derecho is highly efficient when sending large messages, the DDS layering is limited by coordination overheads that soar for small messages (a few KBs).  To sustain a high bandwidth, the entire stack must be optimized around the process of sending and receiving messages. Even microsecond delays significantly impact performance.  Here, we focus on RDMA but the same observation and optimizations would also apply to other high-speed networking technologies (Derecho supports many kinds of networks, including TCP).

We identify two major issues caused by the composition of application threads, DDS API and underlying RDMA library.  First, the library must discover new messages from the application or from the NIC, which is done using a dedicated polling thread.  Even brief delays can have an amplified impact on performance.   Message discovery grows even more complex if a system has multiple incoming or outgoing message streams, as in a DDS.   Second, protocol control messages such as low-level receipt or stability acknowledgments can be surprisingly expensive: The latency to send minimal-sized acknowledgments with RDMA is about the same as the latency to send multiple KBs of application data.

These observations lead to a series of insights.  The first concerns batching. While batching is pervasive, batching at RDMA speeds is not trivial: There is an inevitable tension between not batching sufficiently many events versus waiting too long to accumulate a batch. We present a unifying approach that covers all the stages of the protocol and adapts automatically to the many possible runtime system states.  The approach covers batching not just of messages but also acknowledgments, reducing overheads in both the data and control planes.  Moreover, our technique makes the system far more tolerant of small scheduling delays.

A second insight leads to a modification to the virtual synchrony protocol~\cite{virtual-synchrony} used to deliver messages in identical order.  To maximize speed, Derecho employs a round-robin message delivery order. This favors a steady streaming pattern of communication, but introduces delays when a node is unable to send continuously. Sender delay sensitivity has received little prior attention, but at RDMA speeds becomes a significant issue. We introduce a new null-send mechanism that automatically sends nulls from a lagging sender with minimal delay and without much overhead (it quiesces if the entire application stops streaming messages).
Finally, we show that by restructuring the polling code in a manner that leverages the monotonicity of underlying control data, locks can be released before an RDMA operation is posted. This enables the sender threads to acquire those locks so that preparation of new messages can be done concurrently with the posting of the prior RDMA write.

Prior work on optimizing RPC-style and one-to-one streaming applications for RDMA was highly influential for developers of RDMA systems dominated by one-to-one interactions~\cite{kalia-usenix-atc}.  However, the issues we identify and address were not observed because they arise from the layering of applications over middleware over RDMA, and in some cases are triggered by background events that cause delays.  Similar structures are also seen in message queuing systems, key-value stores that replicate data, atomic multicast and persistent logging. The dramatic speedups Spindle enabled for Derecho and the avionics DDS layered over it thus point to a much broader need, and opportunity.  

\section{Derecho Background}
\subsection{Derecho atomic multicast protocol}
\label{sec:derecho}
We begin with a description of  Derecho, which is more fully described and proved correct in~\cite{derecho-tocs}.  
Derecho manages application membership in a top-level group in accordance with a model called virtual synchrony~\cite{virtual-synchrony}.
In this model, each group evolves through a sequence of views using partition-free state machine replication.
A view change or reconfiguration occurs on failures, node joins and leaves.  These are assumed to be relatively infrequent events.

Derecho models application components as subgroups whose memberships are subsets of the top-level group membership. An atomic multicast API permits any subset of the members to send messages to the entire subgroup. All such messages are delivered safely to the application by every member in the same order\footnote{Derecho atomic multicast is equivalent to Vertical Paxos, and its persistent atomic multicast is equivalent to the classical durable Paxos. However, whereas Paxos uses quorum updates, Derecho delivers every update to every subgroup member, hence every replica holds the complete state.}. Within any subgroup, the developer can designate the members that will initiate atomic multicasts (called senders); this
is done at the beginning of each view and remains fixed until a view change occurs. Messages are then delivered on a round-by-round basis where in each round, one message from every sender is delivered in the order the senders appear in the subgroup-membership list (this property will be  important in Section ~\ref{sec:null_sends_discussion}).
A message is safe to deliver at a node only when the atomic multicast protocol has confirmed that every subgroup member has received it. Messages that are underway when a failure occurs are either delivered to all subgroup members or cleaned up (so that none delivers the message) and then resent in the next membership view.  Messages from external processes are automatically relayed in a failure-atomic manner.

For atomic multicast of small messages, Derecho's key components are the control layer, which runs on a shared state table (SST), a data layer called the small-message multicast (SMC), and a polling framework using predicates that orchestrates the two.

\subsection{SST}
\label{sec:sst_bg}
Derecho's SST (Shared State Table) models each node's local state as a fixed set of monotonic state variables: counters that steadily increase, booleans that shift from false to true, and lists of integers that are updated only via appends or prefix-truncation. These variables are organized into a replicated table, where each node ``owns'' one row, and the columns correspond to the variables.  A node can update its own row, but can only read the other rows.
This policy enables the SST to use a lock-free (hence asynchronous) implementation, in which a node not only specifies the updates it wishes to make, but also controls
when the row is copied (``pushed'') to the other replicas.  RDMA offers a memory fence guarantee that proves useful in designing SST-based protocols: if two updates are initiated
sequentially in different RDMA operations, any application that sees the second update will also see the first.

Although every node in the top-level group maintains the SST, updates pertaining to a single subgroup at a node are only pushed to the other subgroup members. When membership is stable the cost of an update is thus determined by the size of the subgroup, not the size of the entire application.  

In Table~\ref{tab:sst} we see an example SST, used by Derecho for atomic multicasts in a single subgroup. In this example, we have five application nodes with ids $\{0, 1, 2, 3, 4\}$ organized in three subgroups with memberships $\{0, 1, 2\}, \{0, 1, 3\}$ and $\{0, 2, 4\}$. There are two state variables, $received\_num$ and $delivered\_num$ for each subgroup, abbreviated as literals $r$ and $d$ in the table. Messages from each node in a subgroup are received by all members in FIFO order. Thus every message in the subgroup can be assigned a unique sequence number, $seq\_num$ which is its index in the delivery order. The value of $received\_num$ for a subgroup member is the highest $seq\_num\ s$ such that it has received all messages with $seq\_num\leq s$ in the delivery order. Similarly, the value of $delivered\_num$ for a subgroup member is the sequence number of the latest message it has delivered. Both counters are monotonic, starting from $-1$.  

\begin{table*}[t]
  \begin{subtable}{0.36\textwidth}
    \centering
    \begin{tabular}{c | c | c | c ? c | c | c |} 
      \cline{2-7}
      & {\color{blue} r[0]} & {\color{blue} r[1]} & {\color{blue} r[2]} & {\color{red} d[0]} & {\color{red} d[1]} & {\color{red} d[2]} \\
      \cline{2-7}
      {\color{teal} node 0} & 8 & 25 & -1 & 6 & 21 & -1 \\
      \cline{2-7}
      {\color{teal} node 1} & 9 & 21 & --- & 6 & 20 & --- \\
      \cline{2-7}
      {\color{teal} node 2} & 6 & --- & -1 & 6 & --- & -1 \\
      \cline{2-7}
      {\color{teal} node 3} & --- & 23 & --- & --- & 21 & --- \\
      \cline{2-7}
      {\color{teal} node 4} & --- & --- & -1 & --- & --- & -1 \\
      \cline{2-7}
    \end{tabular}
    \captionsetup[subtable]{oneside,margin={2cm,0cm}}
    \vspace*{0.2in}
    \caption{State for atomic multicast}
    \label{tab:sst}
  \end{subtable}
  \hspace*{0.4in}
  \begin{subtable}{0.54\textwidth}
    \centering
    \begin{tabular}{| c | c | c ? c | c ? c |} 
      \cline{1-6}
      {\color{magenta} s[0][0]} & {\color{magenta} s[0][1]} & {\color{magenta} s[0][2]} & {\color{magenta} s[1][0]} & {\color{magenta} s[1][1]} & {\color{magenta} s[2][0]} \\
      \cline{1-6}
      $\{\ldots\}$, 1 & $\{\ldots\}$, 0 & $\{\ldots\}$, 0 & $\{\ldots\}$, 7 & $\{\ldots\}$, 6 & $\{\ldots\}$, -1 \\
      \cline{1-6}
      $\{\ldots\}$, 0 & $\{\ldots\}$, 0 & $\{\ldots\}$, 0 & $\{\ldots\}$, 7 & $\{\ldots\}$, 6 & --- \\
      \cline{1-6}
      $\{\ldots\}$, 0 & $\{\ldots\}$, 0 & $\{\ldots\}$, 0 & --- & --- & $\{\ldots\}$, -1 \\
      \cline{1-6}
      --- & --- & --- & --- & --- & --- \\
      \cline{1-6}
      --- & --- & --- & --- & --- & $\{\ldots\}$, -1 \\
      \cline{1-6}
    \end{tabular}
    \vspace*{0.27in}
    \caption{State for SMC data. \{\ldots\} is a substitute for message content}
    \label{tab:smc}
  \end{subtable}
  \vspace*{0.15in}
  \caption{Sample SST state at node 0 for 5 application nodes and 3 subgroups}
  \vspace*{-0.15in}
\end{table*}

Derecho's SST is implemented using one-sided RDMA writes. Each node maintains a copy of the entire table in its memory. A node updates its state by modifying its row in the local copy and then pushing the updates to the other subgroup members using RDMA writes. As a consequence, a node can read the state of other members of a subgroup directly in its local copy (this is faster than using one-sided RDMA read because in many cases, the data will not have been changed since it was last read). 

Earlier, we noted that the SST is designed for {\em monotonic} data.  For basic types, such as counters, each entry will fit in a cache line.  This maps nicely to RDMA, which is cache-line atomic and sequentially consistent.  
As a result, every subgroup member is certain to
see an increasing sequence of values for every table entry.  For example, when node $0$ sees that $received\_num[1]$ for node 1 increases from 21 to 25, it can conclude that node 1 received the next four messages.  For updates to a list that spans multiple cache lines, SST updates the list data, pushes the update with a first RDMA operation,
then updates a {\em guard:} a monotonic counter used to signal that the data is ready, and pushes it with a second RDMA operation.  The RDMA memory-fencing guarantee ensures that any member that sees the counter update value will also see the updated version of the guarded data.  Notice that all of this is lock-free.

\subsection{SMC}
\label{sec:smc_bg}
SMC (small-message multicast) is a ring-buffer multicast implemented on the SST. Each subgroup has a fixed, configurable number $w$ (for window size) of columns in the SST where each column entry for a particular node is a slot for sending messages in that subgroup. A slot is composed of a message area of a configurable, but fixed size (thus the maximum message size is fixed) and a counter. To send a message from a subgroup member, the application obtains a slot in its row from SMC, generates the message in it and calls send. SMC then updates the slot counter and issues RDMA writes to push the message and the counter to the subgroup members. The slots are utilized in ring buffer order for  consecutive messages. An increase in the value of the counter indicates the presence of a new message - each subgroup member monitors the counter of one slot for each sender in which it expects to receive a message. Messages remain buffered until they have been delivered to the application by every recipient. Thus a sending node needs to track deliveries to know when it can reuse a slot (failing to do so could cause an undelivered message to be overwritten). The intent is that value of $w$  be large enough so that before running out of slots, some slots will have been cleared, enabling continuous sending.

An example of SST columns corresponding to the SMC state are shown in Table~\ref{tab:smc}, where slots are abbreviated using the literal $s$. The first three slots are for subgroup 0, the next two are for subgroup 1 and the last one is for subgroup 2. Thus in node 0's copy of the SST, the counter value of slot[0][0] being 1 for node 0's row indicates that node 0 in subgroup 0 has received 2 different messages in slot 0 from itself, while the counter value of slot[2][0] being -1 for node 4's row indicates that node 0 in subgroup 2 has not received any message from node 4. Only nodes 0 and 1 are senders in subgroup 1, thus the slots in node 3's row are not used. If node 0 were to send a new message in subgroup 1, it will use slot[1][1] of node 0 which will result in the increment of the slot's counter value to 7. The window sizes of 3, 2, and 1 respectively are just to illustrate the concept.  A $w$ value in the range 50 to 1000 would be typical for small messages.

Both SST and SMC guarantee that the memory layout of the application during a view remains unchanged. Thus the required memory can be allocated at each node at the beginning of the view, registered with the NIC and the addresses exchanged with all nodes for RDMA operations.

\subsection{Monotonic predicates over the SST}
\label{sec:preds_bg}
The core of Derecho is its polling thread.  This single thread  evaluates a series of if statements: predicates that test data in the SST, and then a body that will be executed if the predicate is true.  We say that a predicate is monotonic if, when it becomes true for some value $k$, it also holds for values $\leq k$.
Under steady load, Derecho discovers properties such as stability through monotonic predicate evaluation, and then delivers messages in batches.  In contrast, the predicate thread
quiesces when there is no message traffic at all (node A, updating its own state, discovers from the SST that node B is quiescent.  Accordingly, A rings B's doorbell, which wakes it up again).
Thus in the active state, the performance of the predicate thread is central to the performance of Derecho.  Three predicates are of special note:

\textbf{Send predicate}: Detects that the application has prepared new messages that are ready to send.  

\textbf{Receive predicate}: Monitors the SMC slot counter for every sender in the subgroup. When the counter increments, a new message is present, and the receive trigger runs.
The trigger can then increment $received\_num$ (optionally, depending on whether it has increased),
 then push the updated value to other subgroup members.

\textbf{Delivery predicate}: Checks to see if the next message in the delivery order, say with $seq\_num=s$, has become deliverable by checking if every member of the subgroup has received that message ($received\_num[i] >= s,\ \forall\ i$). The trigger delivers the message, updates the receiver's SST row, and then pushes the update.

All three predicates need to run at high speeds without thread scheduling delays.
This explains the decision to employ just a single predicate thread even though many subgroups share the SST: if each had its own predicate thread, they would contend for access to the SST memory as well as internal data structures shared across all subgroups, resulting in locking and possible cache-coherency delays. With a single thread, we lose the opportunity of multi-threaded parallelism, but also eliminate these overheads. Experiments made it clear that with our batching techniques, a single thread can efficiently handle tens of subgroups.

To illustrate these predicates in action, consider a new send by the application thread. The application first acquires a free SMC slot (meaning, the  $delivered\_num$ entry of all subgroup members exceeds that of the slot).
It  constructs a message in the slot and updates the associated counter.  The send predicate detects that the message is ready and initiates RDMA writes to other subgroup members. A lock is needed because the underlying data structures are shared with the predicate thread, and also because multiple application threads may be sending simultaneously.  On the receive side, we see a similar stack, but now the receive predicate senses the incoming messages and the delivery predicate senses that they have become stable and can be delivered.

\section{Spindle optimizations}
We evaluated a baseline version of Derecho's protocols.  Although the system is extremely fast for larger messages~\cite{derecho-tocs}, performance for message sizes and sending patterns typical of a DDS (messages of up to a few KBs, a few dozen topics with subgroups that are heavily overlapping) was low. 
We set out to identify the issues that resulted in such poor numbers. In this section, we detail some of the issues and describe the Spindle techniques that respond to them.

\subsection{In-Place Message Construction and Delivery}
Derecho supports in-place message construction: the application layer supplies a message-constructor as a lambda, and Derecho upcalls to it when a send slot becomes available.  On delivery, messages are passed to the application via pointers.  Spindle leverages this model to reduce communication costs when source and receiver hardware are compatible.  A standard OMG marshaller is used if a setting requires full generality.

\begin{figure}[t]
  \centering
  \includegraphics[width=0.48\textwidth]{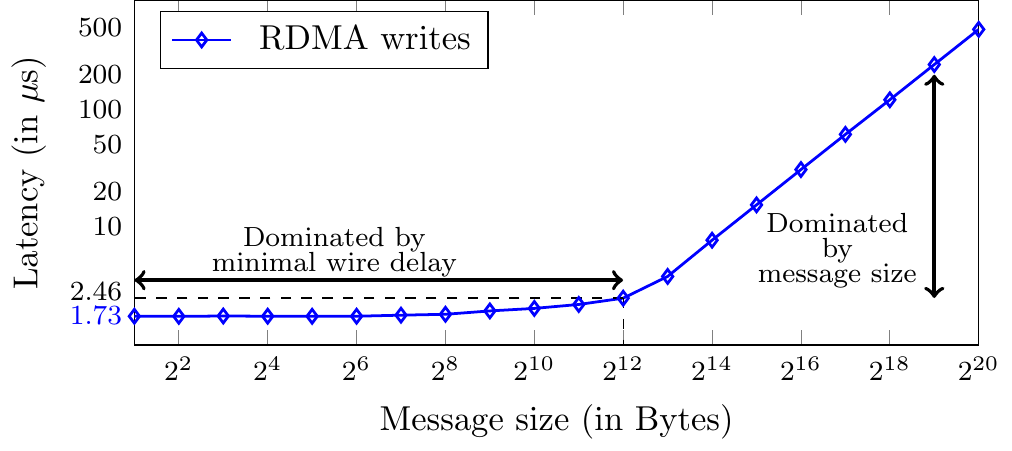}
  \caption{RDMA latency vs data size. Latency is nearly constant for up to 4KB message size.}
  \label{fig:rdma_times}
\end{figure}

\subsection{Opportunistic batching}
One of the main reasons performance of the baseline implementation is low only for small messages is because the latency of sending control data (acks for receiving a message, delivering a message) is comparable to the latency of sending the application messages themselves. Figure~\ref{fig:rdma_times} plots RDMA write latency for different message sizes. Latency for small messages does not increase appreciably with the data size, increasing only marginally from 1.73$\mu s$ for 1-byte data to 2.46$\mu s$ for 4KB data.

The predicates described in Section~\ref{sec:preds_bg} generate an ack for every new message receive and delivery. This turns out to be expensive not only because of the comparatively high latency of control messages as described earlier, but also because posting an RDMA request to the NIC takes \textasciitilde 1$\mu s$ in our setting which means that sending too many acks significantly impacts the efficiency of the predicate thread. We find that the predicate thread in the baseline implementation spends more than 30\% of its time posting  RDMA writes.  

A natural and effective way to address this is to batch events at different stages of the delivery pipeline: send, receive and delivery.   Underlying this observation is Derecho's use of monotonicity: for instance, if 10 messages in a sequence are received before acknowledgment happens, the corresponding $received\_num$ entry can be simply advanced by 10 and a single RDMA write operation issued to push the acknowledgment through the SST. Batching acknowledgments will drastically reduce the number of RDMA writes issued which in turn reduces time spent by the predicate thread posting them.

The usual benefits of batching apply here as well. Batching can improve predicate thread efficiency (by improving locality of predicate evaluation) and can also allow a slow node to catch up with the rest by processing larger batches. In cases with multiple application subgroups, the original protocol makes no distinction between the predicates for different subgroups; that is, the predicate thread evaluates predicates of all the subgroups fairly. When some of the subgroups are not sending messages actively, this reduces the efficiency of the predicate thread, lowering performance. Batching, if done correctly, can help mitigate this issue by adapting batch sizes to the workload. Sending multiple application messages carries an additional benefit, that of sending a larger amount of application data in a single RDMA write, which results in better latency scaling as seen in Figure~\ref{fig:rdma_times}.

Batching is nothing new, but traditionally used fixed batch sizes are ill-matched to RDMA.  If a system ever pauses to accumulate the next batch, the associated delay in sending proves to be remarkably disruptive. In one experiment, we explored waiting to send a fixed batch of messages on top of receive and delivery batching.  Performance collapsed and latency soared even for very small batch sizes.
\begin{figure*}[h]
  \centering
  \includegraphics[width=0.75\textwidth]{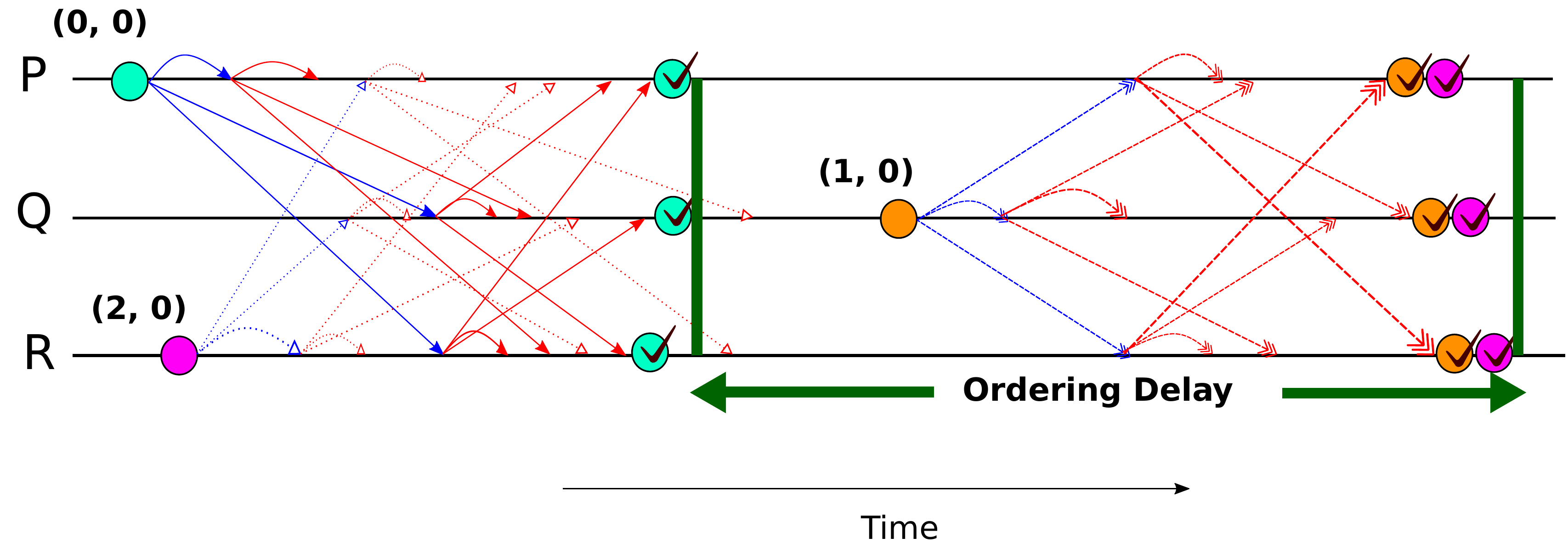}
  \caption{Delays at a sender can impact performance of the system by slowing down other senders}
  \label{fig:null_send_illustration}
\end{figure*}

Accordingly, we created a novel {\em opportunistic multicast batching} technique that covers all stages of atomic multicast. The application thread generates  messages as usual but the send predicate checks to see if multiple messages are ready.  If so it aggregates them on the fly, and sends a batch. The receiver predicate looks through the sequence of slots for each sender, receiving all new messages that it can find. The delivery predicate delivers all messages that have become deliverable, in the right order. Opportunistic batching is {\em self-balancing}: a batch can be smaller or larger depending on the number of events a predicate discovers as it loops. This makes execution more robust to delays by allowing lagging nodes to catch up and does not involve waiting of any kind.

To implement these ideas, we modified the predicates described in Section~\ref{sec:preds_bg} as follows:

\textbf{Send predicate}: The new version of predicate  issues RDMA writes that send all the queued data generated in contiguous ring buffer slots to the other members. If the queued sends have wrapped around the ring buffer, it issues two RDMA writes per remote member accordingly. Since the messages go into discrete slots, each of a fixed size, the predicate pushes the leftover space in the slots too (if messages do not take up the entire slot area). We do not anticipate any downsides to doing so, since the latency for small messages does not rise appreciably and batching allows us to send multiple messages in a single RDMA write.

\textbf{Receive predicate}: For every sender, this predicate goes through the corresponding slots to find all messages that have arrived, stopping at the first empty slot. The trigger updates the $received\_num$ appropriately and pushes the updated value to the other subgroup members.

\textbf{Delivery predicate}: This predicate takes the minimum of the $received\_num$ column for the subgroup members to find all undelivered messages that have been received by all members in the subgroup. Those messages now become deliverable. The trigger delivers all those messages, updates the receiver's SST row, and then pushes the update.

Our changes required a trivial change to the logic for generating application multicasts.  The new version obtains buffers from Derecho, generates messages in them and queues the sends, but does not initiate SMC remote sends. The send predicate does the rest.

It is interesting to contrast our approach with the one-to-one methodology of Kalia et al.~\cite{kalia-usenix-atc}. While they also explore an opportunistic batching technique, their solution will detect at most one client request at a time.  Each request results in a separate RDMA transaction, albeit one crafted to minimize CPU-generated memory-mapped I/Os (MMIOs) by taking advantage of small key and value sizes that fit within an RDMA {\em immediate data} field.  The one-to-one approach is simpler because the setting targeted by Kalia was simpler; in our more complex layered environment, a more complex approach is unavoidable.

\subsection{Null-sends}
\label{sec:null_sends_discussion}
Our second optimization is a novel enhancement to the virtual synchrony model employed by Derecho. Virtual synchrony assumes failures are rare (Paxos is optimized for frequent timeouts and failures, but studies have shown that within a single datacenter, services rarely experience such problems).   We aim to improve performance during the failure-free runs, called epochs. Each epoch has a fixed, ordered membership, known to every member of the system. This way, the order in which messages should be delivered in the epochs does not require consensus among the members; messages are delivered in a pre-determined order which is a function of the group membership. Derecho delivers messages in a round-by-round basis where in each round, one message from every sender is delivered in the membership order.

The core issue we consider is that application sending rates can be variable - it is impossible to guarantee that all senders send messages at a high, steady rate.  Thus, even if the list of senders is identified by the developer, and even if those senders are designed to send data continuously, delays can be introduced by the OS (many systems use IRQ balancing to spread interrupt-handling overheads among the available cores), by lightweight thread scheduling, or because a thread must wait for a lock held by the predicate thread (Section~\ref{sec:thread_synchronization}).  In one experiment we even saw a situation in which a library used a C++ spin-lock rather than a mutex lock.  The intent was that spin-locks would be faster than a mutex, but in fact the experiment revealed a case where this was exceptionally slow: the C++ 17 implementation of spin-locks turns out to be unfair on NUMA hardware and can favor one thread while disadvantaging other threads.  In the particular case, Linux decided which core each thread would run on, and the application's sender thread turned out to be running very slowly compared to the Derecho thread that checked for new messages!

The null-messages optimization can't really fix these sources of application delay, but it does reduce their global impact.  When a sender is not prepared to send its next message, messages from other senders may be delayed in delivery, owing to Derecho's fixed round-robin message-delivery ordering (Section ~\ref{sec:derecho}). We illustrate this issue in Figure~\ref{fig:null_send_illustration}. We have three nodes, P, Q, R, that each send a message denoted by green, orange and pink circles, respectively. The blue arrows denote the send of a message, while the red arrows denote the send of an acknowledgement. We see that sender Q sends its message much later than P and R. Since the delivery order is P, Q, R, we see that while P's message is delivered as soon as each node learns that it has been received by all the nodes, R's message has to wait until Q's delayed message is delivered. With Derecho's ring buffer implementation multiple messages can be sent at the same time, but delays in sending by a single member can leave multiple messages stuck waiting at the receivers, if that delayed sender is next in the round-robin order. The ring buffers of active senders will soon fill up with undelivered messages, preventing them from sending more messages.

The obvious way to deal with this issue is to detect when a sender has fallen behind and then send dummy $0$-sized messages (called {\em nulls}) from that sender to expedite the delivery of application messages from other senders. At the time of delivery, null messages can simply be discarded and the resulting sequence of delivered messages will still be same across the members. The problem  is that we detect the potential delay in receiver logic, and yet a null-send is logically a sender-side action.  Moreover, designing an efficient null-send scheme that decides when and how many nulls to send at RDMA speeds has not been explored in the literature: Prior null-send protocols ran on older TCP networks, where the processor was so fast relative to the network that small sending delays did not risk appreciable performance loss.

Sending a null too soon at RDMA speeds and latencies interferes with normal message delivery, because the sender may have been about to send a legitimate application message. Over time, this will result in sending too many nulls which, owing to RDMA's relatively high 1-Byte latency (Figure~\ref{fig:rdma_times}), will add up to a significant cost. On the other hand, sending a needed null even a few microseconds too late is undesirable because these tiny delays still represent significant lost bandwidth.  We desire four properties:
\begin{enumerate}[leftmargin=*]
\itemsep0em
\item \textbf{Sender-invariance}: Performance with only a subset of senders sending continuously does not drop appreciably.
\item \textbf{Low-overhead}: Performance does not degrade significantly when all senders are sending actively compared to the same case without any null-send scheme in-place.
\item \textbf{Correctness}: Under all circumstances of senders sending at different rates (and possibly some senders not sending at all), the delivery pipeline never stalls.
\item \textbf{Quiescence}: When all senders are inactive, the system attains a quiescent network state where no nulls are sent.
\end{enumerate}

Spindle's null-send scheme is simple to describe but challenging to prove correct: When a sender node receives a message, it sends a single null message if this null message will precede the received message in the delivery order. This is done by checking the per-sender index of messages sent and global sequence numbers, and thus efficient.

Correctness and Quiescence can be proved as follows.
Denote a message in a subgroup as $M(i, k)$ where $i$ is the sender rank (in the senders list) and $k$ is the sender index, equal to the number of messages it has sent in the subgroup. Round robin delivery imposes a total ordering $<$ on the messages: $M(i_1,k_1) < M(i_2, k_2) \iff k_1 < k_2 || (k_1 = k_2 \wedge i_1 < i_2)$.
Assume that node $i$ receives a message from a sender with rank $j$ in round $k, M(j, k)$. Without loss of generality, assume $i < j$. The null-send scheme will send a null iff current round number of sender $i$, $l$ (equal to the number of messages sent by $i$), is such that $l < k$.

Suppose a null is sent. It is an easy induction to deduce that $l=k-1$ (consider what happened when $i$ received $M(j, k-1)$). That is to say, that the null-send keeps every sender within one round of each other in terms of the number of messages it has received vs. sent. Thus after $M(j, k)$ has been received by all nodes, their own round number is greater than or equal to $k$. This statement is imprecise for nodes with rank $> j$ but without loss of generality, we can take $j$ to be the highest ranked sender. This implies that sends of all messages that precede $M(j, k)$ have been initiated, meaning that $M(j, k)$ will be delivered barring failures. Hence no deadlock arises.

We now show that the system reaches a quiescent state when no application messages are being sent. The complicating factor is that $M(j, k)$ may itself be null. However, if $M(j, k)$ is null, it was sent by sender $j$ in response to another message $> M(j,k)$ received by it. This chain cannot go on forever and thus will finally terminate in a non-null, application message. Thus, if no sender is actively sending messages, no nulls will be sent. This explains why we do not need to check if the received message is a null. In fact, sending a null in response to another null may expedite the delivery of subsequent application messages (this would be useful if messages arrive in different orders at different nodes).
It is straightforward to combine null-sends with batching: After the receiver predicate finishes an iteration, it sends the determined number of nulls as a single integer. 

The intention is that Spindle's null-send scheme should dynamically adjust to real-time delays, lagging nodes, and other disruptions, maintaining high levels of performance when a sender is unintentionally delayed. If an application deliberately will not send messages from some source for an extended period of time, it should declare the number of rounds of inactivity to the subgroup members which can then appropriately modify the message delivery sequence. If a node is never going to send again, it can be marked as a non-sender when it joins the subgroup or later, during a reconfiguration. Thus, the (small) overheads of the null-send scheme are seen only in the event of unanticipated delay.

\subsection{Efficient thread synchronization}
\label{sec:thread_synchronization}
Efficient thread synchronization is crucial to high performance. For Derecho, we noted one potential inefficiency: when application threads interact with the polling thread, a lock is required that protects against concurrency conflicts, yet can delay sending.  To assess the degree to which such a problem arises, we looked deeply at the structure of Derecho's predicates and the code blocks they trigger. What we found is that many predicates interleave access to SST data with RDMA write operations. RDMA writes are costly to post: they can consume 20-50\% of the time spent in the entire predicated test and code block.  
Accordingly, we restructured all Derecho predicates so as to place the RDMA write calls only at the end. This can be done safely as a predicate's logic does not depend on the state of the SST at a remote node, but only what is present in the local SST. Then we can safely release the lock before we proceed to issuing RDMA writes.  Any parallel access of the SST by other threads is safe because of the following two properties of the SST: (1) simultaneous reads and writes to the SST are safe since the variables fit within cache-lines and (2) any updates to the variables being pushed that might occur between when the predicate releases the lock and when the push actually occurs are monotonic.  Thus, the eventual push will simply batch the original information with additional data.

\subsection{Delays caused by the receiver}
Derecho offers safe and consistent delivery of messages. The application ``acts on'' (or consumes) a message only when it is delivered. The protocol delivers messages in the critical path - the predicate thread discovers that a message is deliverable, calls into the application with that message, and updates and pushes the corresponding $delivered\_num$ after the upcall returns. Waiting until all receivers have consumed the message makes it safe for Derecho to reuse the associated slot for sending a fresh message.

As a result, delays in the delivery upcall have a dramatic performance impact, because the predicate thread cannot continue its run until the upcall returns. To quantify this effect, we built an application in which message delivery upcalls take 1$\mu s$, 100$\mu s$ or 1$ms$ and found that performance decreases by about 9\%, 90\%, and 99\% on average, respectively. For larger delays of 100$\mu s$ and 1$ms$, performance degenerates to one message delivered per delay time. This finding confirms that the protocol is highly sensitive to the time taken by the application in processing the message. 

To mitigate the impact of delays in delivery, we offer two viable options: (1) Applications can support a batched delivery upcall, which consumes all messages that are deliverable. If processing a batch of delivered messages takes less time overall, we obtain performance speedups. (2) Applications can simply move the data into a separate memory area via memcpy and return from the upcall. For small messages, costs of memcpy are not terribly high. We evaluate the overhead of memcpy during delivery in Section~\ref{sec:memcpy_eval}.

\section{Evaluation}
In this section, we evaluate the impact of the Spindle optimizations discussed in the previous section. The evaluation focuses on throughput, defined as the amount of application data delivered per unit time (GB/s, averaged over all nodes). Each test is run 5 times - we plot the average values and show error bars corresponding to one standard deviation. We test on our local cluster consisting of 16 machines connected with a 12.5GB/s (100Gbps) RDMA Infiniband switch. Each machine has 16 physical cores and 100GB of RAM.

We evaluate a multitude of scenarios with one or multiple subgroups with one or all of them sending messages actively, senders sending continuously or with delays, delays in various parts of the protocol. For each optimization, we show the cases most directly impacted by the change.  Subsequent optimizations are evaluated on top of the previous optimizations, showing incremental improvements.  Finally, we look at the overall impact of Spindle on the application that motivated our effort:  an avionics DDS.

\subsection{Opportunistic batching}
\subsubsection{Single subgroup continuous sending}
Many systems have just one replication group, for example to replicate a component or data or to support event notifications. In this case, all senders continuously stream messages in a tight loop. We vary the subgroup size from 2 to 16 using message sizes 1B, 128B, 1KB and 10KB, in three patterns - all senders (every member is a sender), half senders (only half of the members are senders) and just one sender. Small message sizes can go as far as few hundred KBs, but by limiting it to 10KB, we can leverage the power of aggregation while keeping within the limit. Consistent with the SMR approach, all members are receivers in all cases and deliver all sent messages in the same order. Each sender sends a total of 1~million messages. The experiment finishes when all messages have been delivered.

\begin{figure}[t]
  \centering
  \includegraphics[width=0.48\textwidth, height=4.5cm]{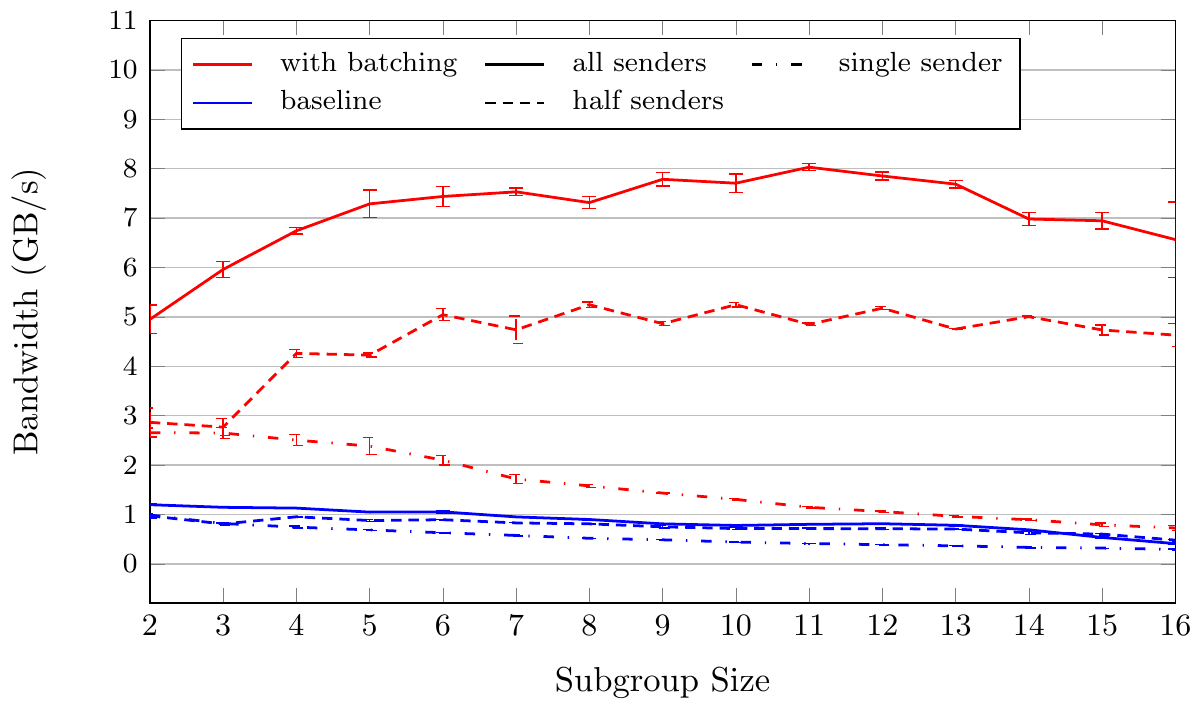}
  \caption{Performance for single subgroup with opportunistic batching. Performance improves by up to 16X.}
  \label{fig:batching_main}
\end{figure}

Figure~\ref{fig:batching_main} plots performance for this test for 10KB messages and compares it against the baseline performance. As is clear from the graph, opportunistic batching alone outperforms the baseline by about 9X for all senders, 6X for half senders and 3X for one sender on average. The peak bandwidth attained is 8.03GB/s for 11 members, giving a maximum network utilization of 64.2\%. Performance also scales much better with increasing number of senders, for instance, it is 16X of the baseline performance with 16 senders. Performance with just one sender declines with the subgroup size as the algorithm pays the price of increased coordination overheads.

\begin{figure}[t]
 \centering
 \includegraphics[width=0.48\textwidth, height=4.5cm]{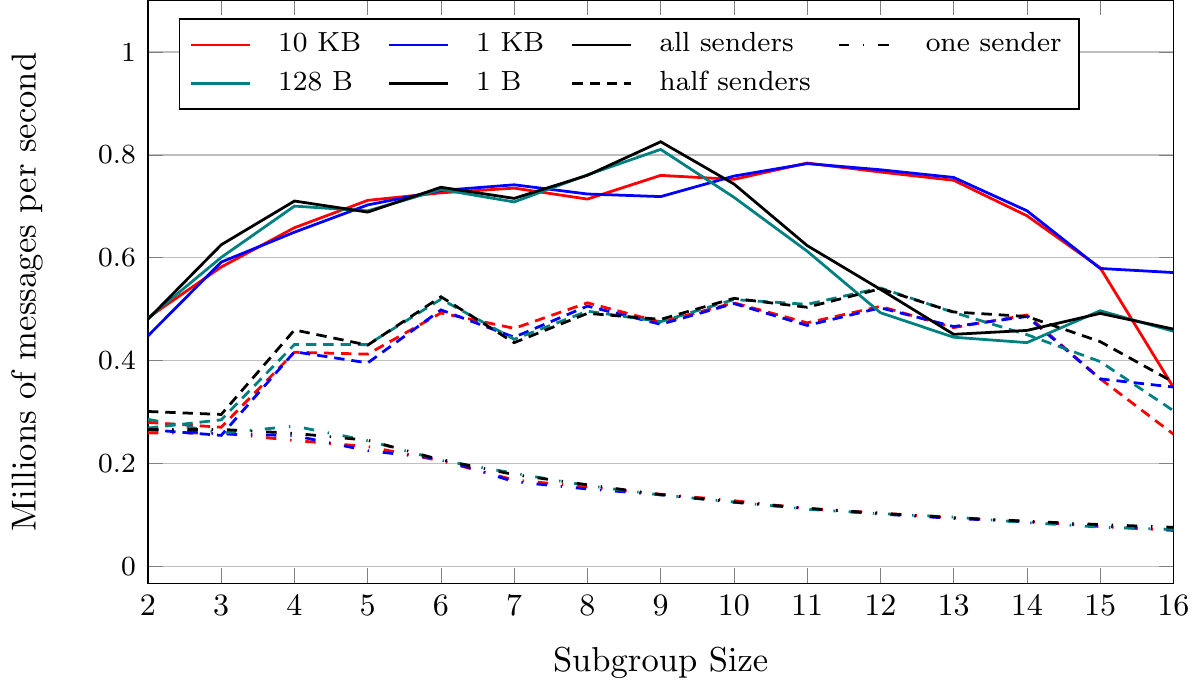}
 \caption{Rate of delivery for single subgroup with opportunistic batching.  Derecho has a second communication larger, RDMC, for very large subgroups or messages.  Although RDMC was not evaluated in our work, shifting to it might be advisable for subgroups with more than 12 members.}
 \label{fig:batching_main_msgs}
\end{figure}

Consistent with Figure~\ref{fig:rdma_times}, the performance is proportional to the data size for both the baseline and the optimized version. In other words, the number of messages delivered per second remains about the same for different small message sizes. Figure~\ref{fig:batching_main_msgs} confirms this observation for the optimized version. 
As such, all subsequent experiments only show data for the 10KB case.

\begin{figure}[t]
  \centering
  \includegraphics[width=0.48\textwidth, height=4.5cm]{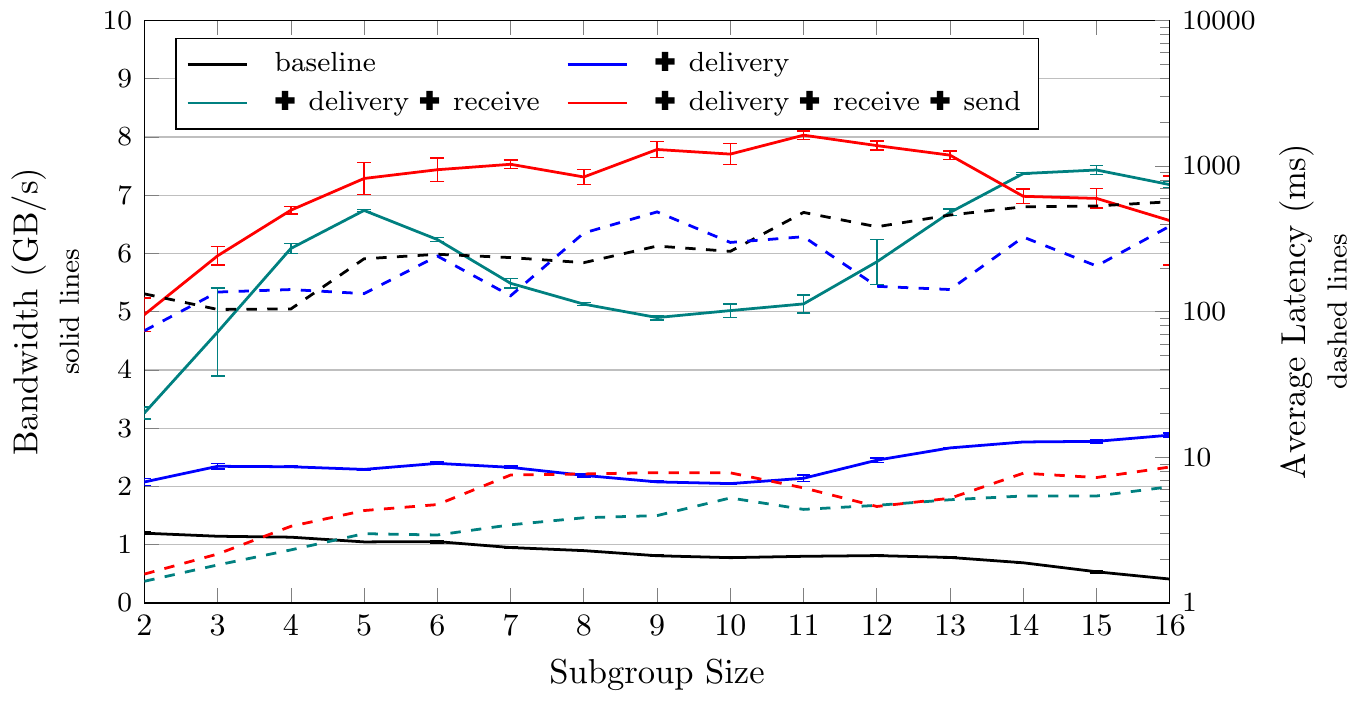}
  \caption{Performance gains with batching applied to successively more stages of the pipeline for all senders.  Throughput (left Y-axis) is shown by solid lines, and latency (right Y-axis) using dashed lines.  Both metrics show significant improvements relative to our baseline system.  }
  \label{fig:batching_incremental}
\end{figure}

It is interesting to learn the impact of batching at different stages of the protocol. Figure~\ref{fig:batching_incremental} shows the incremental effect of adding delivery, receive and send batching successively.  It is particularly noteworthy that our optimizations improve {\em both} throughput and latency across the full range of subgroup sizes.  In contrast, as noted earlier, traditional forms of sender-side batching sharply increase latencies, and may significantly reduce bandwidth by leaving the RDMA network idle while waiting to accumulate the next batch.

We computed some metrics for the 16 senders case to gain an insight into the improvements. Comparing the baseline against the optimized version, we find that the number of RDMA write requests goes down from 18.2M to 1.1M, time spent by the polling thread in posting RDMA writes goes down from 64.84s to 4.29s and the sender thread spends time waiting to find a free buffer only for 52.7\% of the much reduced experiment time (as opposed to 97.6\% of the total runtime of the baseline).

\subsubsection{Suitable ring buffer size}
Batch sizes for our batching optimization are influenced by the subgroup window size. After all, the number of messages that can be sent or received in one batch is limited by the number of slots. An unreasonably small window size does not allow for optimal batching, while an excessively large window size forces the predicate thread to cover too large a memory area. In this experiment, we measure the performance of the single subgroup all senders case, varying ring buffer size.

\begin{figure}[t]
  \centering
  \includegraphics[width=0.48\textwidth, height=4.5cm]{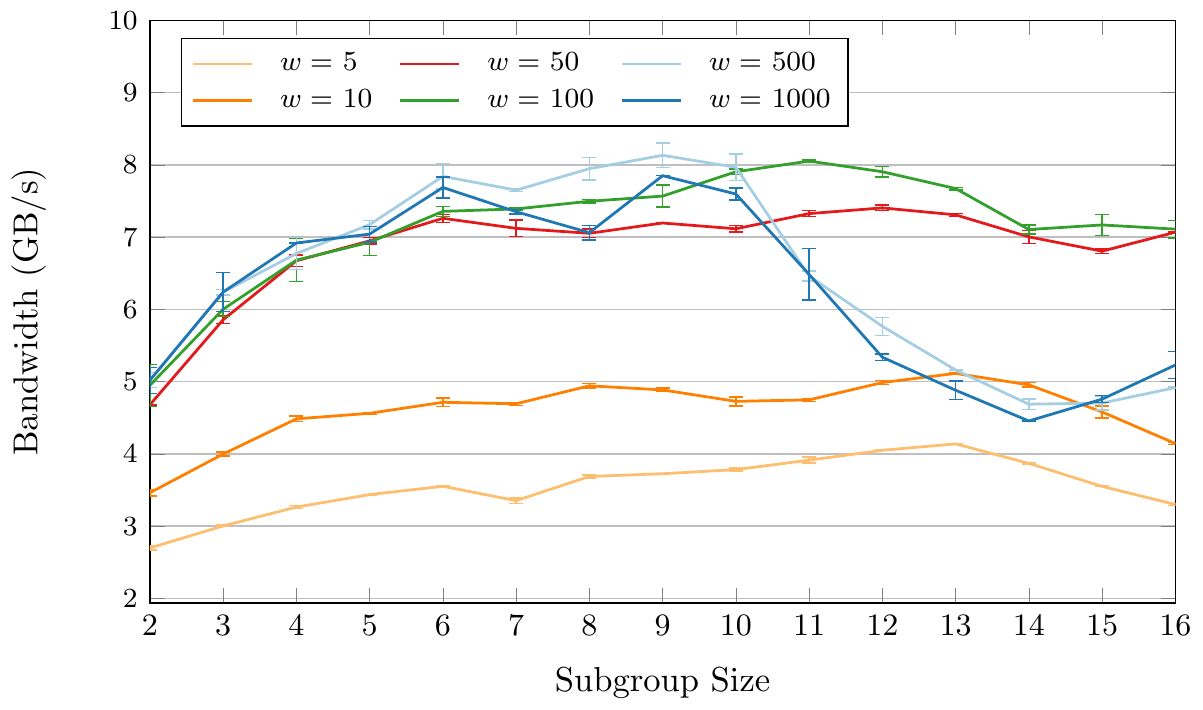}
  \caption{Performance with different window sizes when all nodes are sending messages continuously}
  \label{fig:batching_window_size}
\end{figure}

Figure~\ref{fig:batching_window_size} plots the results. Even a small window size of 5 increases performance by 4.5X average compared to the baseline with 100 window size! The highest performance is obtained for a window size of 100. Consequently, all our experiments for 10KB message size use a window size of 100. It is important to note that performance with window sizes of 500 or 1000 starts declining after 10 nodes, quite likely because the polling area increases considerably and large batches of application messages (if 200 messages of 10KB are sent in one RDMA write, total data size is a little less than 2MB) do not give good throughput with a simple multicast send scheme of SMC (sequential send). This suggests that applications should use a window size around 100 instead of pinning large buffers with RDMA.

For the single subgroup case, the SST at each node consists of just two columns for the subgroup state $received\_num$ and $delivered\_num$ which takes 16 bytes of space per row, and the slots for the SMC. The total space for the slots at each node is
\[
  n * w * (m + 8)
\]
where $n$ is the number of nodes (rows), $w$ and $m$ are the window size and maximum message size (8 less than the size of the slot which also contains a counter) for the subgroup. For 16 members, 10KB message size, and $w$ = 100, the total space per subgroup amounts to roughly 16MB. This suggests that applications can easily scale to tens of subgroups with the total memory allocated within few hundred MBs.

\begin{figure*}[t]
  \centering
  \begin{subfigure}[t]{0.25\textwidth}
    \includegraphics[width=\textwidth, height=3cm]{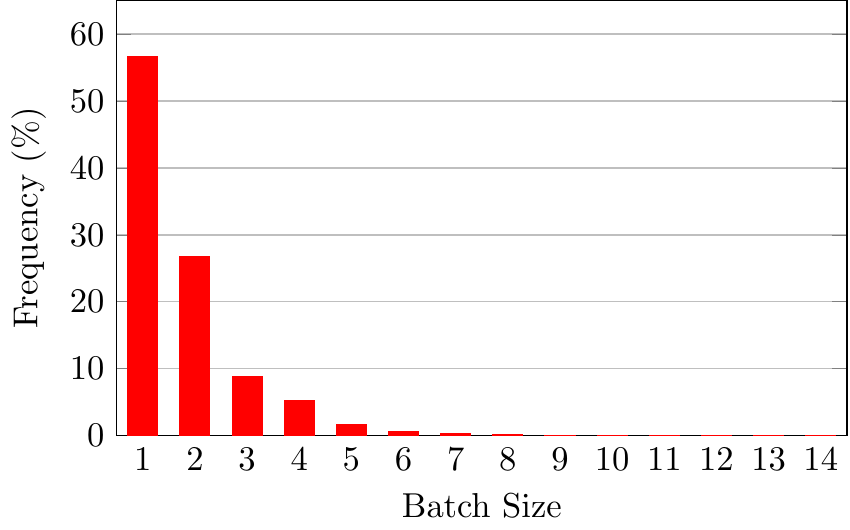}
    \caption{Sends}
    \label{fig:batching_send_histogram}
  \end{subfigure}
  \begin{subfigure}[t]{0.40\textwidth}
    \includegraphics[width=\textwidth, height=4cm]{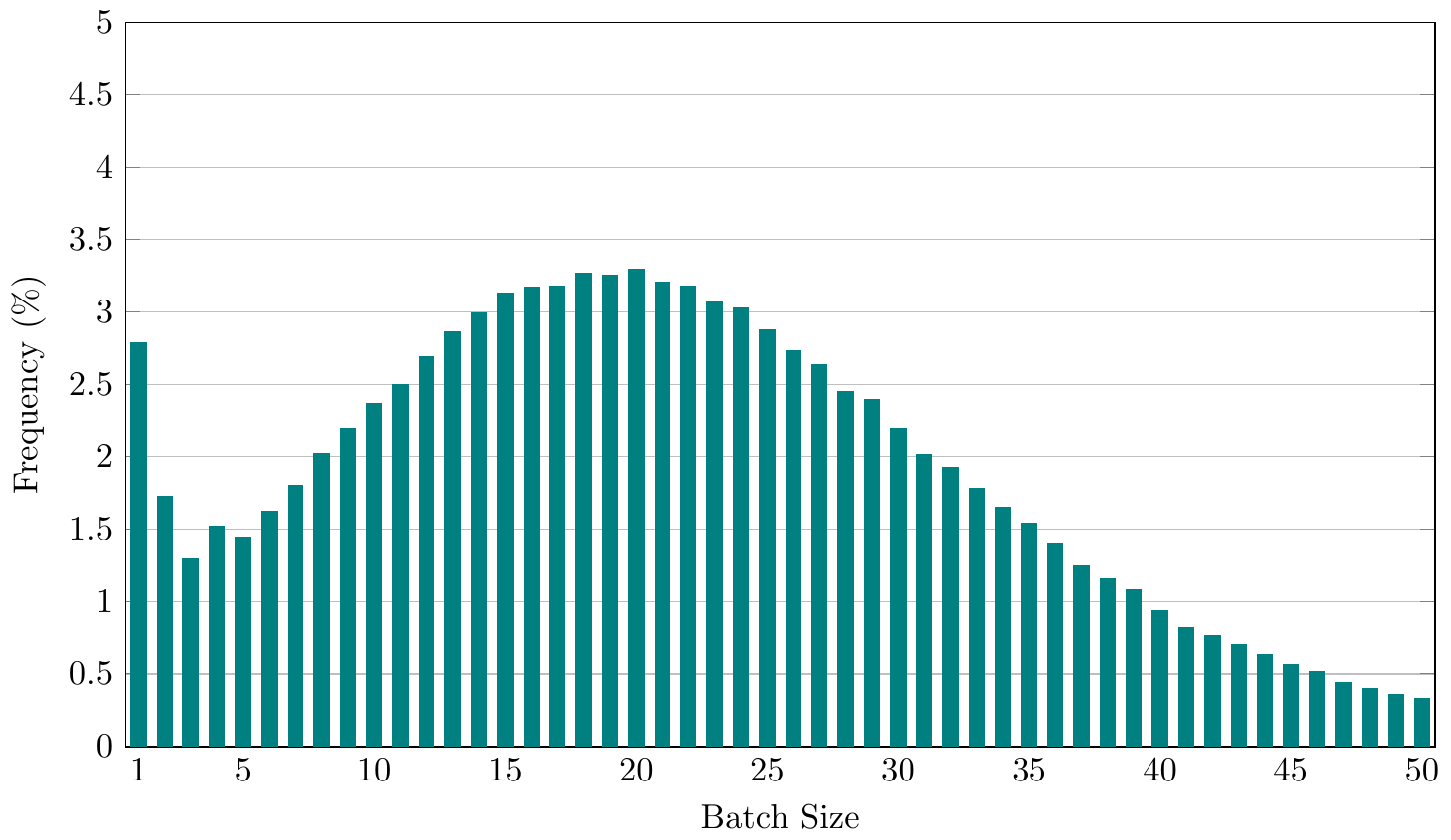}
    \caption{Receives}
    \label{fig:batching_receive_histogram}
  \end{subfigure}
  \begin{subfigure}[t]{0.25\textwidth}
    \includegraphics[width=\textwidth, height=3cm]{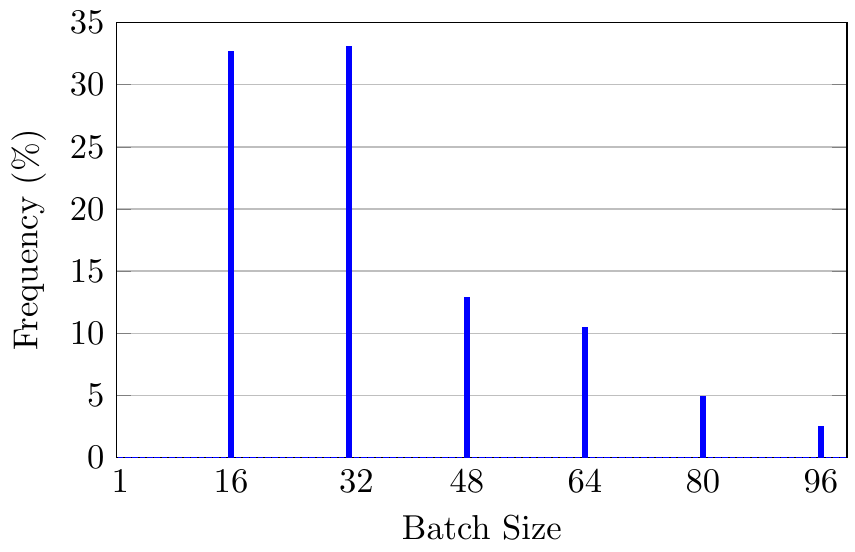}
    \caption{Deliveries}
    \label{fig:batching_delivery_histogram}
  \end{subfigure}  \caption{Batching histograms for the three protocol stages. Receive merges data streams from all senders, forming larger batches. Delivery computes an extra level of stability over all members, forming even larger batches.}
  \label{fig:batching_histograms}
\end{figure*}

It is interesting to learn the batch sizes for different steps of the pipeline. Figure~\ref{fig:batching_histograms} plots the histograms for a window size of 100 for the single subgroup, 16 senders case. Messages are typically sent in small batches of less than 5, while most delivery batches are multiples of 16 suggesting that about 1-5 messages from each sender are typically delivered in a batch. Different mean batch sizes for send, receive, and delivery is further proof that a rigid batching scheme with fixed batch sizes is unlikely to work well in practice, especially in heterogeneous environments where nodes are running at different speeds.

\subsubsection{Single active subgroup}
In this case, all nodes belong to all subgroups, but each node continuously sends 1M messages in just one of them. Our goal here is to expose inefficiencies inherent in the baseline which evaluates all subgroups' predicates fairly and show how opportunistic batching compensates for those inefficiencies.

\begin{figure}[t]
  \centering
  \includegraphics[width=0.48\textwidth, height=4cm]{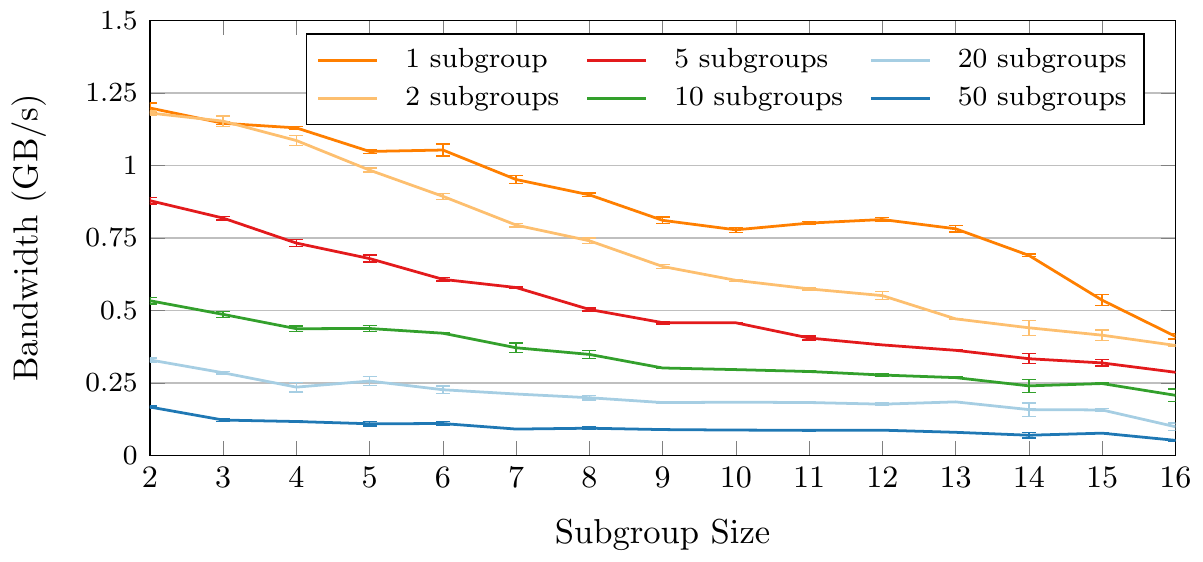}
  \caption{Performance of baseline for single active subgroup}
  \label{fig:baseline_single_active}
\end{figure}

\begin{figure}[t]
  \centering
  \includegraphics[width=0.48\textwidth, height=4.5cm]{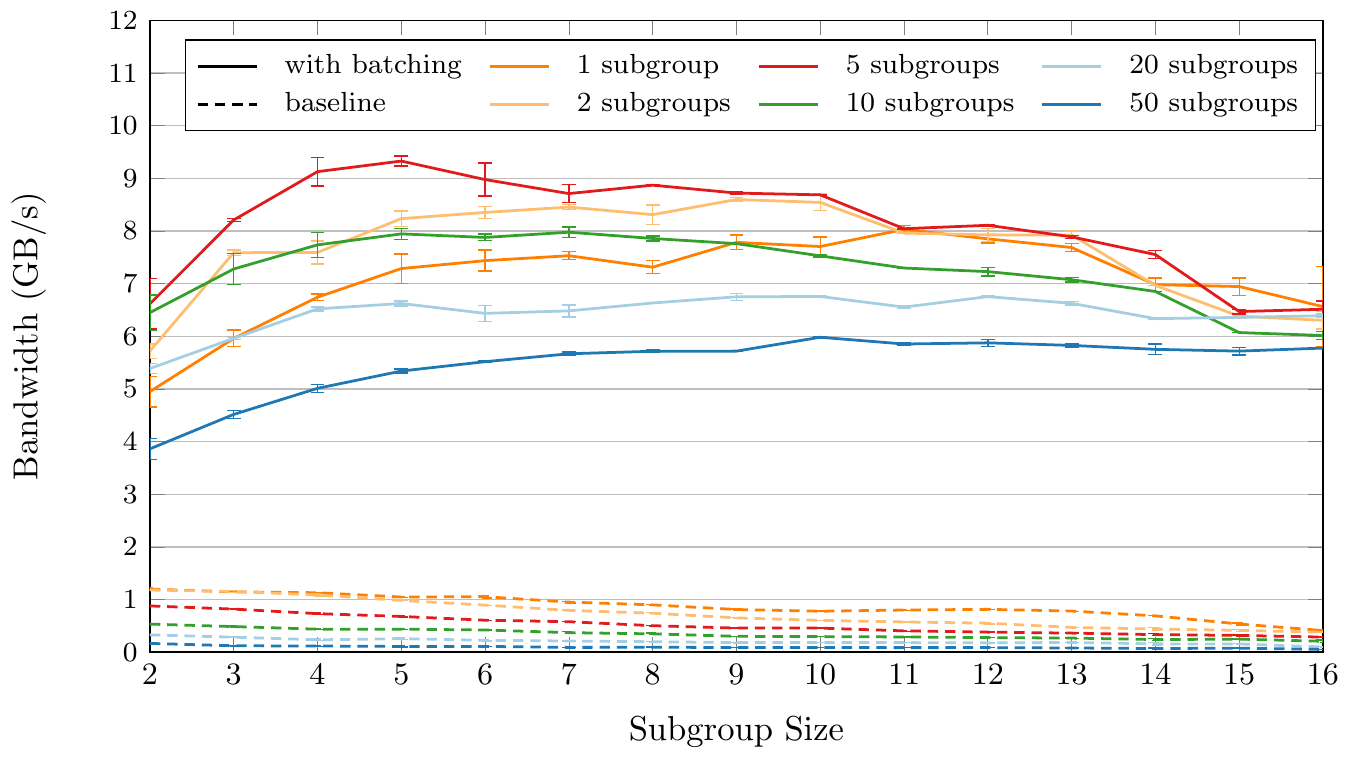}
  \caption{Performance with opportunistic batching for single active subgroup}
  \label{fig:batching_single_active}
\end{figure}

Figure~\ref{fig:baseline_single_active} plots the results for the baseline implementation. As expected, performance consistently decreases with increasing number of subgroups. Even adding a single inactive subgroup degrades performance by 18\% on average, while the performance with 50 subgroups is one-tenth of the performance with the sole active subgroup. On the other hand, Figure~\ref{fig:batching_single_active} plots the results for the optimized version. We see that adding more subgroups does not decrease performance invariably but increases it in some cases! Performance with 5 and 10 subgroups is better than the same with 1 and 2 subgroups, respectively. This is an artifact of batching. Presence of delays can sometimes result in more efficient executions, due to larger average batch sizes. Clearly, there is a lot of potential in a more adaptive batching scheme that adjusts according to the circumstances. Even with 50 subgroups, the performance declines much more graciously compared to the baseline. This stability should help developers feel confident that a decision to use overlapping subgroups will not harm application performance.

For the baseline, for a sample run with 16 nodes, the percentage of time spent evaluating the active subgroup's predicates goes down from 54\% for 2 subgroups to less than 15\% for 50 subgroups. With opportunistic batching, this number is about 99\% for 2 subgroups, 90\% for 10 subgroups and 48\% for 50 subgroups. The average batch sizes for sends, receives and deliveries increase from \{1.72, 22.18, 35.19\} for 1 subgroup (Figure~\ref{fig:batching_histograms}) to {\{6.20, 49.36, 127.74\}}, {\{21.67, 79.15, 334.48\}} and {\{50.45, 207.46, 638.57\}} for 2, 10 and 50 subgroups, respectively. This shows the adaptability of opportunistic batching to real-time delays.

Opportunistic batching also vastly improves performance for the multiple active subgroups case, where multiple subgroups are actively sending messages. However, performance drops considerably with increasing number of subgroups. We infer that the predicate thread spends an increasing amount of time posting RDMA writes for the different subgroups, delaying timely sending of application messages. Our optimization of efficient thread synchronization resolves most of these overheads, hence, we evaluate this case in Section~\ref{subsec:efficient_thread_sync}.

\subsection{Null-send scheme}
\label{sec:nulls_eval}
\subsubsection{Delayed sending}
\begin{figure}[t]
  \centering
  \includegraphics[width=0.48\textwidth, height=4.5cm]{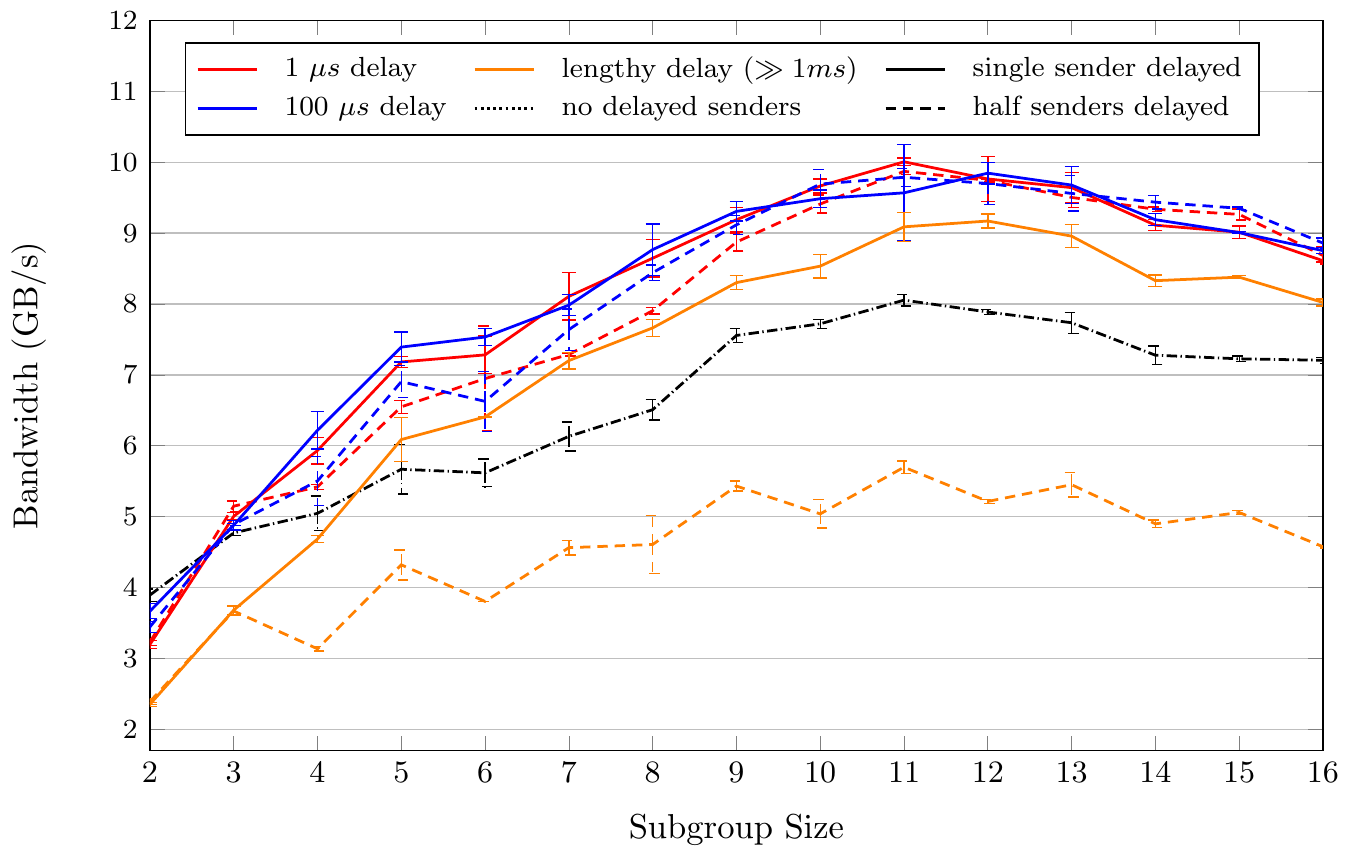}
  \caption{Data for sender delay test with null-sends}
  \label{fig:null_sender_delay}
\end{figure}

In any real system, there may be unpredictable delays in sending. In this experiment, we simulate such a case for the all senders case by introducing a fixed delay after each send at either one or half of the senders. Senders that are not delayed send as fast as possible. We tried several different delays: 1) 1$\mu s$, a minimal delay close to the network latency, 2) 100$\mu s$, much larger than network latency, yet realistic for applications, 3) a lengthy delay. In each case, the delay is implemented with a busy-wait loop. We measure bandwidth after a fixed number of messages have been delivered. As detailed in Section~\ref{sec:null_sends_discussion}, the baseline protocol does nothing to adjust for these kinds of delays. This is the primary test for the null-send scheme.

Figure~\ref{fig:null_sender_delay} plots the results, which are surprising. For every case other than where half senders are delayed indefinitely, performance increases, peaking at 10.0GB/s. This is because small delays lead to larger average batch sizes and large delays lead to more efficient bandwidth utilization by the remaining senders. This shows that the system adapts very well to real-time delays.

In case of 16 nodes sending 1M messages each with 1 sender delayed by 100$\mu s$, the delayed sender sends one or more nulls in 517K iterations of the receiver predicate while a continuous sender sends them in only 189K iterations. The average inter-delivery time between consecutive messages from a continuous sender and a delayed sender comes down drastically and in fact, decreases from 3.779$\mu s$ for 2 nodes to 1.617$\mu s$ for 8 nodes and 1.192$\mu s$ for 16 nodes. This confirms that nulls accelerate delivery of application messages.

\subsubsection{Continuous sending}
Nulls may be inserted even when all senders are sending continuously because of inevitable small relative motion between the members in sending and receiving messages. This could potentially either reduce performance if nulls interfere with application messages or increase performance if nulls compensate for batching-induced delays. In a real setting, where sending patterns are more varied, null-sends will improve performance as established by the previous experiment.

In this experiment, we measure the impact of null-sends when all senders are sending continuously in a subgroup. We compare Derecho with only opportunistic batching against Derecho with null-sends on top of opportunistic batching.

\begin{figure}[t]
  \centering
  \includegraphics[width=0.48\textwidth, height=4.5cm]{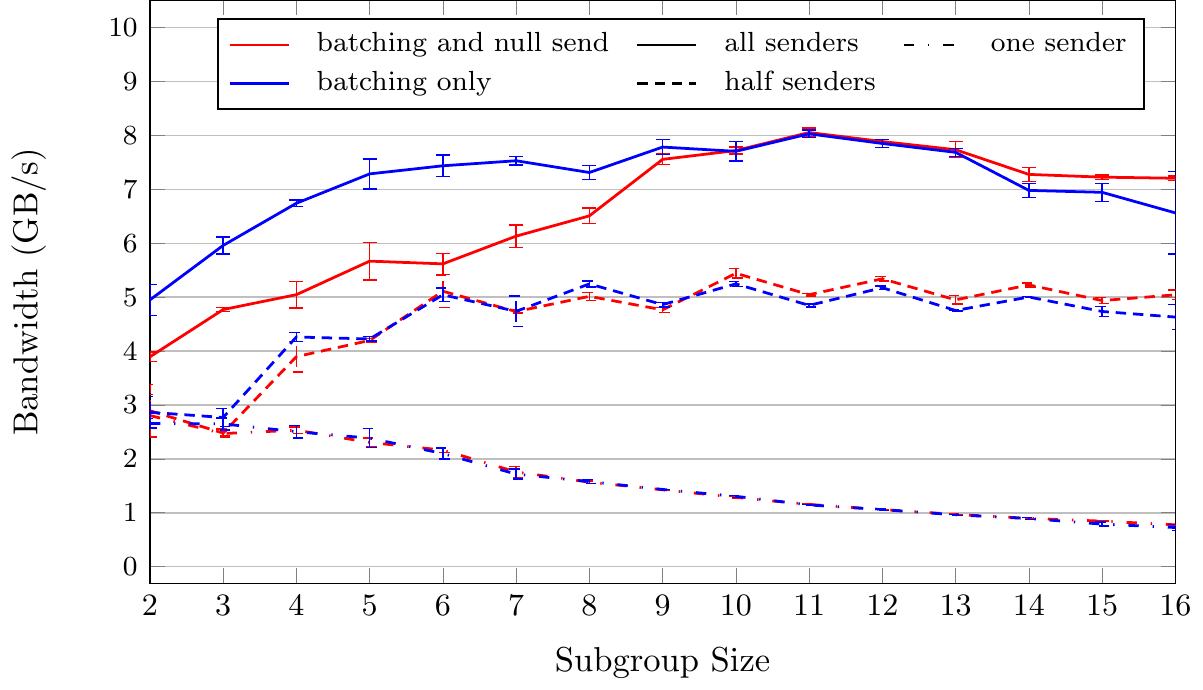}
  \caption{Impact of null-sends on continuous sending}
  \label{fig:null_main_comp}
\end{figure}

Figure~\ref{fig:null_main_comp} compares performance. For all senders, performance is initially worse because there is less scope for improvement, and therefore nulls have a minor but deleterious impact. With larger subgroup sizes, small delays become more prominent. Here null-sends accelerate message delivery leading to improved performance. The drop for smaller nodes is significant for all senders (up to 25\%) and almost negligible for half senders. No nulls can ever be sent for one sender; the graph confirms that no overhead is introduced.

\subsubsection{Additional Null-Send experiments}
We also conducted additional experiments that exposed the null-sending scheme to increasing complex and disruptive delays, such as by declaring all members of a shard as senders, but then having just one member do all the sends.  For reasons of brevity, we omit details, but in all cases the mechanism successfully compensated, allowing the active senders to run at full speed, while filling any gaps caused by inactive senders.  Null-sends are not always the entire solution: in Sec. ~\ref{sec:null_sends_discussion} we mentioned a case in which an unfair C++ spin-lock caused a library to malfunction in a way that drastically slowed some senders.  The null-send mechanism prevents such slowdowns from propagating to other senders, but doesn't fix the slowdown itself.  Still, the resulting pattern highlighted the slow sender.  This focused our attention, and ultimately enabled to track down the root cause, at which point we were able to modify the library in question to use a mutex lock, restoring full performance.

\subsection{Efficient thread synchronization}
\label{subsec:efficient_thread_sync}
We evaluate the effect of restructuring predicates to move RDMA writes to the end and release locks before issuing the writes. We evaluate the performance for the single subgroup, all senders and the multiple active subgroups cases.

\begin{figure}[t]
  \centering
  \includegraphics[width=0.48\textwidth, height=4cm]{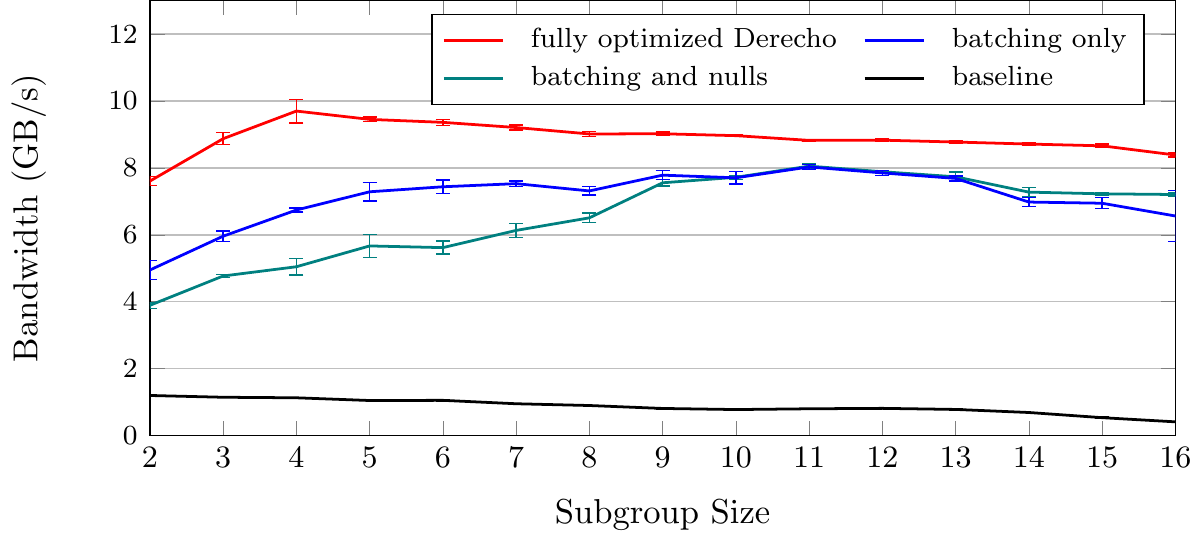}
  \caption{Impact of efficient synchronization}
  \label{fig:efficient_sync}
\end{figure}

\begin{figure}[t]
  \centering
  \includegraphics[width=0.48\textwidth, height=4.5cm]{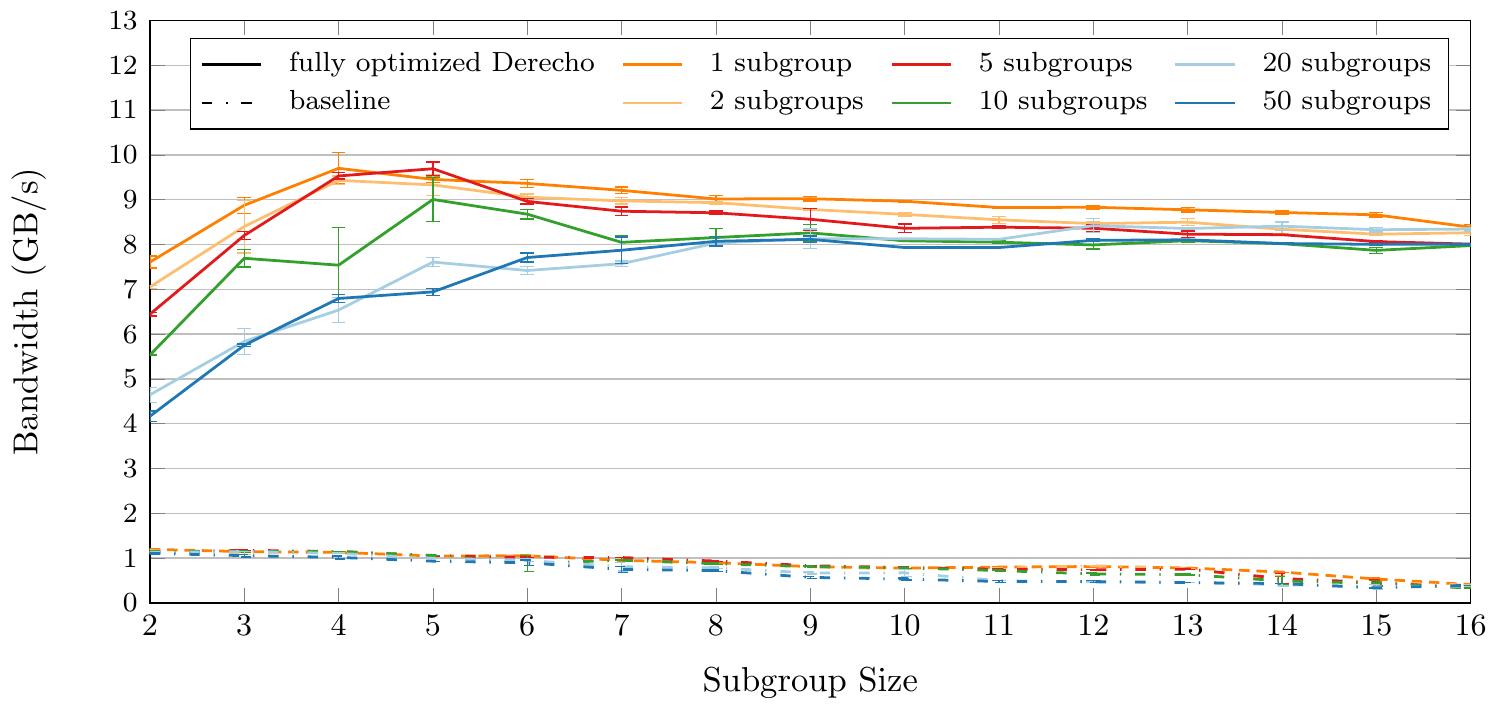}
  \caption{Final performance with all optimizations for multiple active subgroups}
  \label{fig:batching_multiple_active}
\end{figure}

Figure~\ref{fig:efficient_sync} plots the results for single subgroup. The optimization, on top of batching and nulls, improves performance considerably by about 1.4X average. The maximum network utilization of 77.6\% is reached for 4 members which stays very stable all the way to 16 members. 

Figure~\ref{fig:batching_multiple_active} plots the results for multiple active subgroups, comparing them with the baseline. The results show excellent scaling with the number of subgroups. The performance remains relatively stable for all subgroup sizes.

\subsection{Delays caused by memcpy}
\label{sec:memcpy_eval}
\begin{figure}[t]
  \centering
  \includegraphics[width=0.48\textwidth, height=4cm]{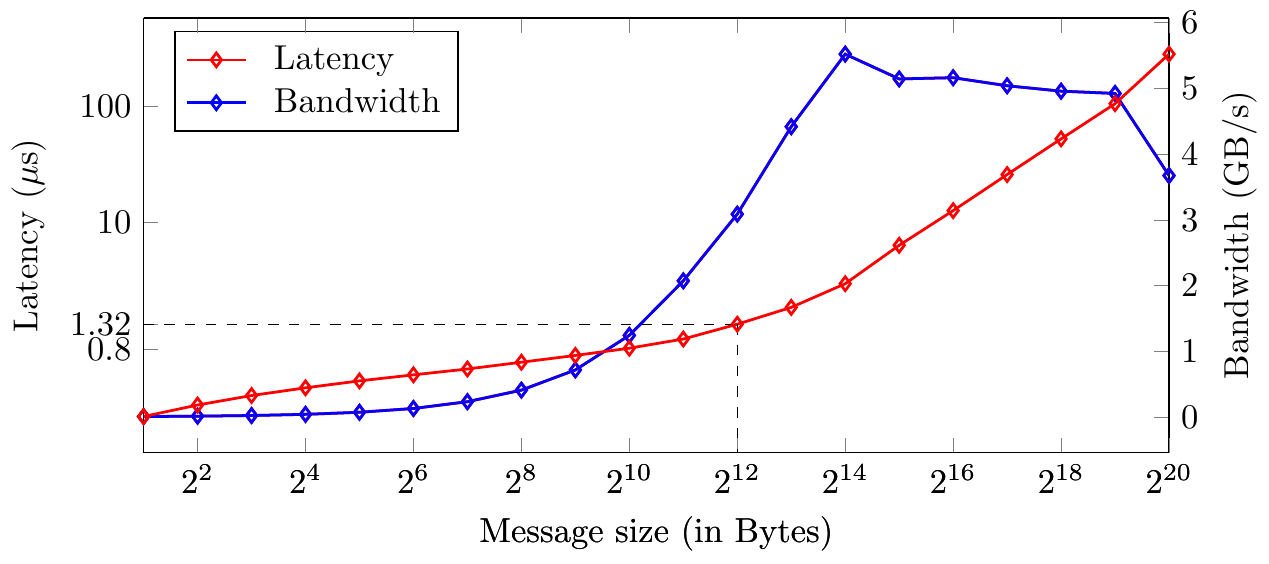}
  \caption{Performance of memcpy with data size}
  \label{fig:memcpy_main}
\end{figure}

RDMA is based on the zero-copy idea: memory copy within a node is much slower than remote copy over a network. High RDMA speeds impose considerable strains on application memory management. In our optimized Derecho implementation, for instance, sends and delivery must finish quickly to stay close to the optimal performance. It may not be practical to avoid memory copy when generating a message in the ring buffers (for example, if the application receives data out of band from external clients) or to give up ownership of a message immediately after delivery. However, memory copy is not that expensive for small messages. Figure~\ref{fig:memcpy_main} measures the latency and bandwidth of memcpy on one of our machines. The latency remains low up to a few KBs, then quickly deteriorates for large message sizes.

For this reason, we evaluate a pragmatic approach where the application copies data from external buffers into Derecho-provided slots before sending and copies data out of the ring buffers in delivery. We again evaluate the single subgroup case for 10KB messages. For smaller messages sizes of 1B and 128B, memcpy carries much less overhead.

\begin{figure}[t]
  \centering
  \includegraphics[width=0.48\textwidth, height=4cm]{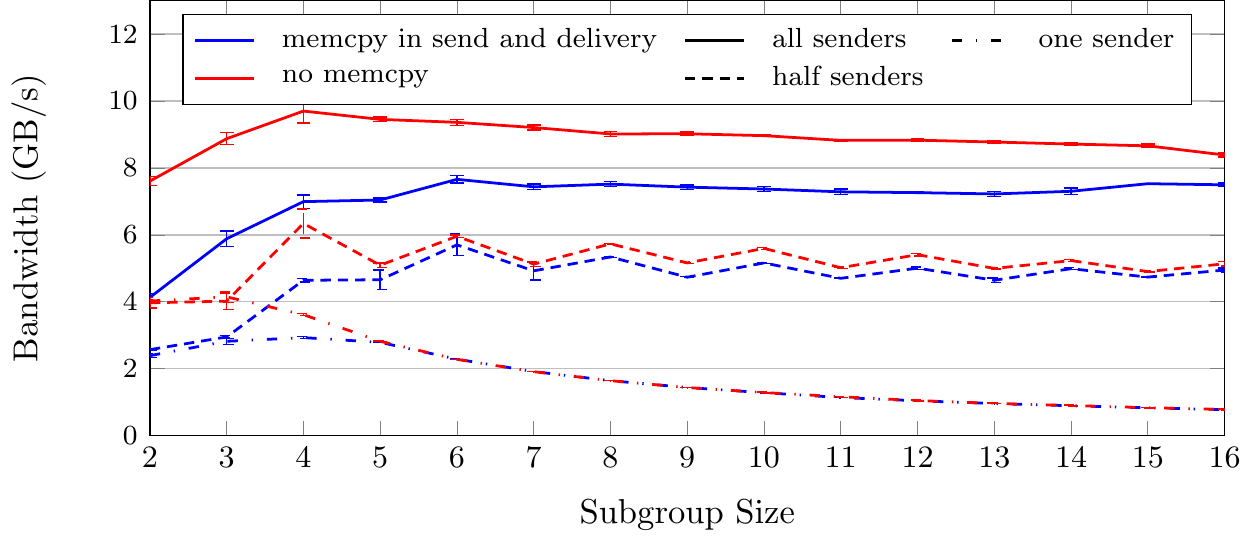}
  \caption{Performance with memcpy in send and delivery}
  \label{fig:memcpy_send_delivery}
\end{figure}

Figure~\ref{fig:memcpy_send_delivery} shows that there is a decline for all senders, though the bandwidth still remains consistently around 7.5GB/s. Performance declines slightly for half senders, while there is almost no decline for the single sender case as the memcpy induced delays are likely absorbed into the coordination overheads. We also evaluated this case for the extreme case of 1B messages and observed no performance loss.

\subsection{Final results}
\begin{figure}[t]
  \centering
  \includegraphics[width=0.48\textwidth, height=4cm]{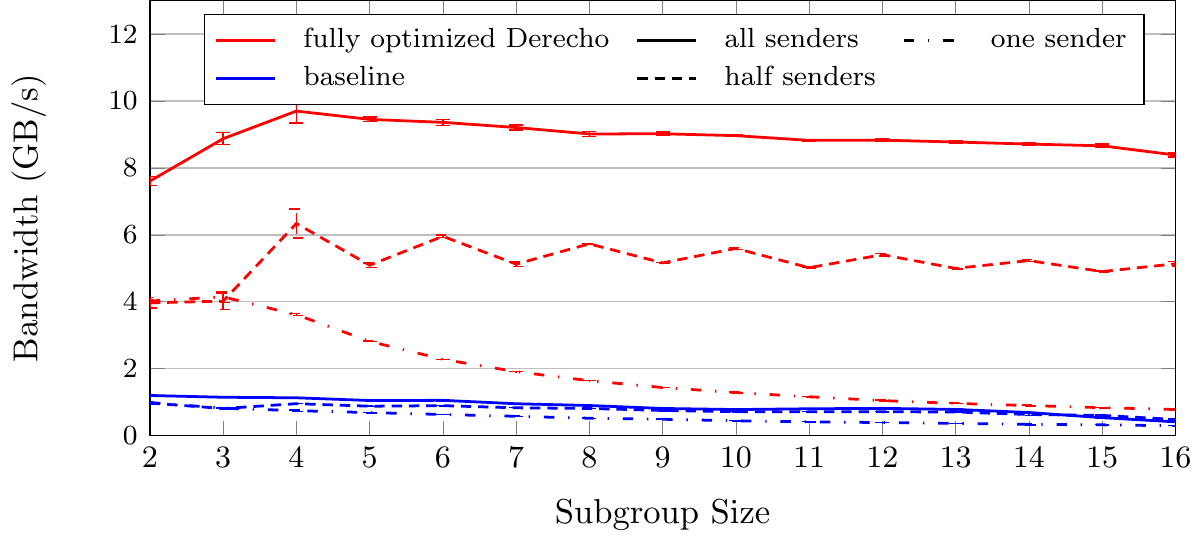}
  \caption{Final throughput numbers for a single subgroup. }
  \label{fig:final_thp}
\end{figure}

\begin{figure}[t]
  \centering
  \includegraphics[width=0.48\textwidth, height=4.5cm]{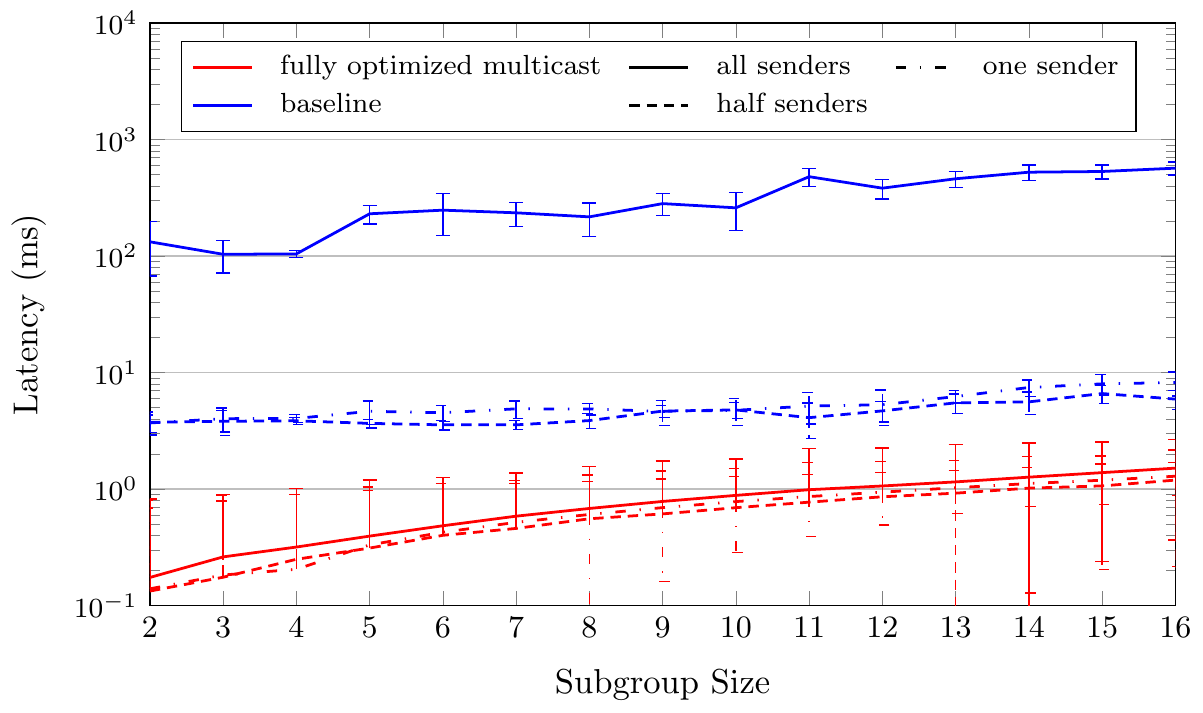}
  \caption{Final latency numbers for a single subgroup}
  \label{fig:final_lat}
\end{figure}

Figure~\ref{fig:final_thp} shows final numbers for the single subgroup case with all senders, half senders and one sender.  As noted, although
our optimizations were focused on throughput, Figures ~\ref{fig:batching_incremental} and ~\ref{fig:final_lat} both show substantial improvement in latency.  Note that the logarithmic Y-axis scale magnifies error bars for small latency values.

\subsection{DDS evaluation on Spindle + Derecho}

Our DDS prototype maps the DDS API,  Data-Centric Publish-Subscribe (DCPS) to the underlying Derecho system by forming a single Derecho 
"top-level" group that includes all publishers and subscribers, then forms subgroups for each topic containing only the processes that publish or subscribe to that topic (the actual Spindle DDS also supports "external clients" that connect to the DDS via TCP or RDMA, requiring an extra relaying step, but we did not evaluate that mode of use).  The user then defines data types and publish and subscribe topics (abstractly, a Global Data Space or GDS).   Importantly, the Spindle DDS permits developers to construct messages "in place", and then mark them as ready to send.  Had we used a model in which the application allocates space elsewhere to create its messages, the resulting overheads would have sharply reduced performance.  

We tested performance for a single DDS topic with a single publisher and varying number of subscribers. We defined a {\em Sequence} data type, which represents a simple byte sequence, to be exchanged among the entities.  Note that because the data type did not require serialization, our experiment does not encounter the potentially significant delays that such a step would have introduced. The publisher continuously publishes in that topic 1 million samples of type Sequence, each of 10KB size. To stress the network performance, publishers and subscribers are all on different nodes.

An OMG DDS can offer various QoS levels. Our DDS has four: 1. \textbf{Unordered}: Data is delivered to the application without waiting for stability and discarded after delivery. This is relevant for applications that do not need any kind of ordering or reliability. 2. \textbf{Atomic multicast}: This maps directly to Derecho's atomic multicast, and data is discarded after the delivery upcall. 3. \textbf{Volatile storage}: Incoming data is copied and saved on the receiver node's memory (this allows a joining subscriber to catch up). 4. \textbf{Logged storage}: Data is additionally appended to a log file on SSD storage, and is used for debugging, and in applications that track the evolution of a reported measurement over time.  

\begin{figure}[t]
  \centering
  \includegraphics[width=0.48\textwidth, height=4.5cm]{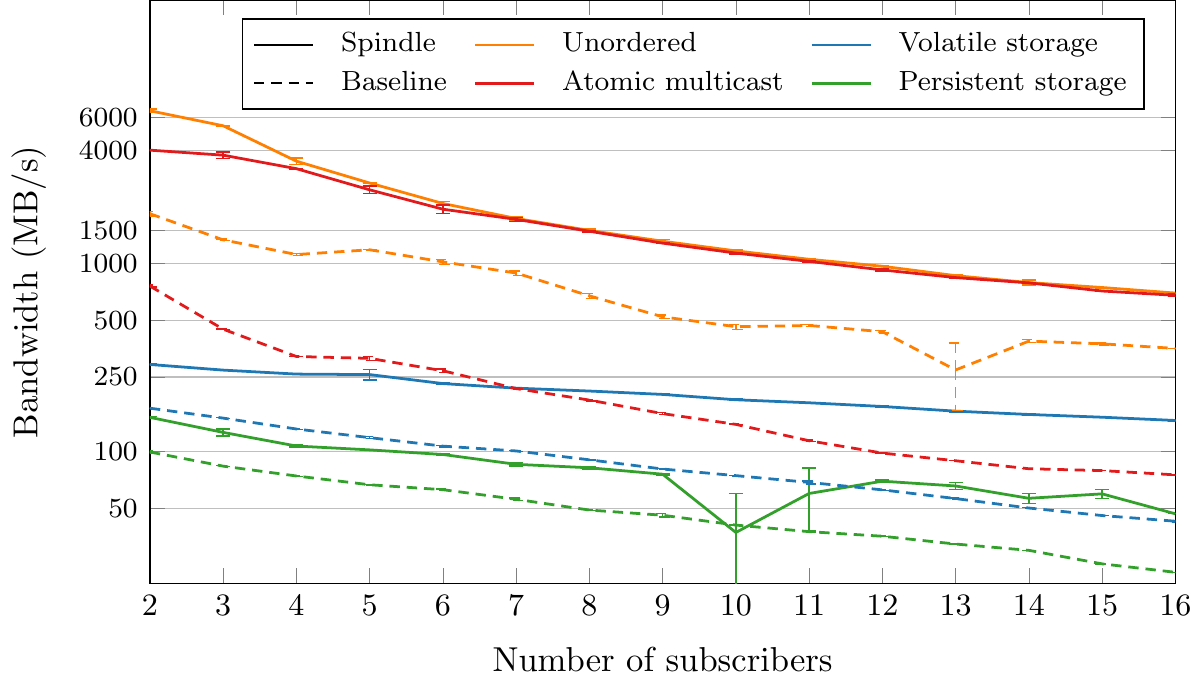}
  \caption{DDS performance improvements with Spindle optimizations for all 4 QoS levels}
  \label{fig:dds_bw}
\end{figure}

Figure 18 compares bandwidth for our baseline DDS implementation with one that uses the Spindle optimizations. We see that Spindle improves performance for all four cases relative to the baseline. Whereas Spindle-DDS has nearly the same performance for unordered and atomic multicast mode, notice that the pre-Spindle baseline's performance decreases considerably with each additional QoS level. This validates our effort to reduce communication overheads.  Interestingly, Spindle's performance improvements even carry over to the volatile and persistent storage modes, despite the fact that these are limited by memory copying and disk I/O costs.  The finding supports our hypothesis that whole stack optimization yields a steadier end-to-end data stream even when a variety of potential bottlenecks are present.  

\section{Related Work}
\label{sec:related-work}
Although our work was motivated by an avionics DDS layered over Derecho, there has been other work on using RDMA to accelerate
state machine replication~\cite{APUPaxos, Dare-SMR, derecho-tocs} and key-value storage~\cite{farm-rdma, herd-rdma, FaSST, kalia-usenix-atc}.   The separation of control from the data plane originated in BarrelFish and Arrakis~\cite{Barrelfish, Arrakis} and the iX $\mu$-kernel~\cite{iX} employs a similar separation of layers.  Earlier we compared Spindle with the work of Kalia et. al. ~\cite{kalia-usenix-atc}, but there are also interesting similarities to $\mu${T}une~\cite{asriraman18}, a thread-level coordination package for  low-latency, high-throughput gRPC-based $\mu$-services.
One of the Spindle optimizations involves sending null messages to avoid delays if a sender is not ready to send a new multicast when its turn arises.
This idea was first explored in the Totem~\cite{Totem} and Transis~\cite{Transis} systems, and similar mechanisms have been used in modern Ring Paxos protocols~\cite{Quema,jalilimarandi2010ring}.  
\section{Conclusion}
\label{sec:conclusion} 
We reported on Spindle, a  methodology for optimizing complex, multi-participant middleware to leverage RDMA.   Beyond the avionics DDS scenario that motivated our work, the Spindle techniques would also be applicable in other coordination-based distributed systems and middleware components running on high-speed networks.

\section{Acknowledgments}
This work was supported by AFRL under the SWEC program, Microsoft, Nvidia/Mellanox and Siemens.

\bibliographystyle{ACM-Reference-Format}
\interlinepenalty=10000
\nocite{*}
\bibliography{references}

\end{document}